\documentclass{article}
\usepackage{spconf}
\usepackage[cal=boondox,%
calscaled=.97,%
bb=ams,%
bbscaled=.94,%
frak=euler,%
frakscaled=.97,%
scr=esstix]{mathalpha}

\usepackage{amsmath}
\usepackage{amsmath}
\usepackage{amsthm, amssymb, amsfonts, mathtools, cases}

\usepackage[%
style = ieee,%
citestyle = numeric-comp,%
bibencoding = utf8,%
datamodel = software,%
backend = biber,%
giveninits = true,%
abbreviate = true,
maxbibnames = 5,%
maxcitenames = 2,%
sortcites=true,%
doi=false,%
isbn=false,%
url=false,%
eprint=false%
]{biblatex}

\usepackage{pgfplots}
\pgfplotsset{compat=newest}
\usepgflibrary{plotmarks}
\usetikzlibrary{
  tikzmark,
  calc,
  arrows,
  decorations.pathmorphing,
  shapes,
  matrix
}

\usepackage{graphicx}
\graphicspath{ {./figs/} }
\DeclareGraphicsExtensions{.png,.pdf}
\usepackage{xcolor}
\usepackage{url}

\definecolor{cadetblue}{rgb}{0.37, 0.62, 0.63}
\definecolor{burntorange}{rgb}{0.8, 0.33, 0.0}
\definecolor{americanrose}{rgb}{1.0, 0.01, 0.24}
\definecolor{applegreen}{rgb}{0.55, 0.71, 0.0}
\definecolor{darkmagenta}{rgb}{0.55, 0.0, 0.55}
\definecolor{peru}{rgb}{0.80, 0.52, 0.25}
\definecolor{navy}{rgb}{0.0, 0.0, 0.5}
\definecolor{maroon}{rgb}{0.5, 0.0, 0.0}
\definecolor{gold}{rgb}{1.0, 0.84, 0.0}
\definecolor{crimson}{rgb}{0.86, 0.08, 0.24}

\usepackage{newfloat}

\usepackage{subcaption}

\usepackage[inline]{enumitem}
\usepackage{etoolbox}

\usepackage{algorithm, algpseudocode}
\makeatletter
\newcommand{\algmargin}{\the\ALG@thistlm}
\makeatother
\newlength{\whilewidth}
\algdef{SE}[parWHILE]{parWhile}{EndparWhile}[1]
  {\parbox[t]{\dimexpr\linewidth-\algmargin}{%
     \hangindent\whilewidth\strut\algorithmicwhile\ #1\ \algorithmicdo\strut}}{\algorithmicend\
   \algorithmicwhile}%
\algnewcommand{\parState}[1]{\State\parbox[t]{\dimexpr\linewidth-\algmargin}{\strut #1\strut}}

\newcommand{\IntegerP}{\mathbb{N}}
\newcommand{\IntegerPP}{\mathbb{N}_*}
\newcommand{\Real}{\mathbb{R}}

\newcommand\given{{\mathbin{}\mid\mathbin{}}}
\newcommand\vect[1]{\mathbf{#1}}
\newcommand\vectgr[1]{\boldsymbol{#1}}

\providecommand\given{} 
\newcommand\SetSymbol[1][]{
  \nonscript\,#1\vert \allowbreak \nonscript\,\mathopen{}}
\DeclarePairedDelimiterX\Set[1]{\lbrace}{\rbrace}%
{ \renewcommand\given{\SetSymbol[\delimsize]} #1 }
\DeclarePairedDelimiterX\innerp[2]{\langle}{\rangle}{#1
  \mathop{}\delimsize\vert\mathop{} #2}
\DeclarePairedDelimiterX\norm[1]\lVert\rVert{\ifblank{#1}{\:\cdot\:}{#1}}

\DeclareMathOperator{\Fix}{Fix}

\DeclareMathOperator{\prox}{Prox}

\DeclareMathOperator{\soft}{Soft}

\DeclareMathOperator{\argmin}{argmin}

\usepackage{thmtools}
\usepackage[unq]{unique}

\declaretheoremstyle[%
headfont=\normalfont\bfseries,
notefont=\mdseries,
notebraces={(}{)},
bodyfont=\normalfont,
postheadspace=1ex
]{mystyle}

\declaretheorem[style=mystyle,
                name=Theorem,
                refname={theorem,theorems},
                Refname={Theorem,Theorems}
]{thm}

\declaretheorem[style=mystyle,
                name=Proposition,
                refname={proposition,propositions},
                Refname={proposition,propositions}
]{prop}

\newlist{thmlist}{enumerate}{1}
\setlist[thmlist]{label=\textbf{(\roman{*})}, ref=\thethm(\roman{*}), noitemsep}

\newlist{lemlist}{enumerate}{1}
\setlist[lemlist]{label=\textbf{(\roman{*})}, ref=\thelemma(\roman{*}), noitemsep}

\newlist{exlist}{enumerate}{1}
\setlist[exlist]{label=\textbf{(\roman{*})}, ref=\theexample(\roman{*}), noitemsep}

\newlist{factlist}{enumerate}{1}
\setlist[factlist]{label=\textbf{(\roman{*})}, ref=\thefact(\roman{*}), noitemsep}

\newlist{proplist}{enumerate}{1}
\setlist[proplist]{label=\textbf{(\roman{*})}, ref=\theprop(\roman{*}), noitemsep}

\newlist{asslist}{enumerate}{1}
\setlist[asslist]{label=\textbf{(\roman{*})},
  ref=\theassumption(\roman{*}), noitemsep}

\newlist{deflist}{enumerate}{1}
\setlist[deflist]{label=\textbf{(\roman{*})}, ref=\thedefinition(\roman{*}), noitemsep}

\newlist{algolist}{enumerate}{1}
\setlist[algolist]{label=\textbf{(\roman{*})}, ref=\thealgo(\roman{*}), noitemsep}

\newlist{claimlist}{enumerate}{1}     
\setlist[claimlist]{label=\textbf{(\roman{*})}, ref=\theclaim(\roman{*}), noitemsep}

\newlist{applist}{enumerate}{1}
\setlist[applist]{label=\textbf{(\roman{*})}, ref=\thesection(\roman{*}), noitemsep}

\newlist{MyEnumSec}{enumerate}{1}
\setlist[MyEnumSec]{label=\textbf{\thesection(\roman{*})},
  ref=Item~\thesection(\roman{*}), noitemsep}

\newlist{MyEnumSubSec}{enumerate}{1}
\setlist[MyEnumSubSec]{label=\textbf{\thesubsection(\roman{*})},
  ref=Item~\thesubsection(\roman{*}), noitemsep, wide = 0pt, leftmargin = *}

\usepackage[capitalize]{cleveref}

\crefname{thm}{Theorem}{Theorems}
\crefname{prop}{Proposition}{Propositions}
\crefname{assumption}{Assumption}{Assumptions}
\crefname{lemma}{Lemma}{Lemmata}
\crefname{definition}{Definition}{Definitions}
\crefname{example}{Example}{Examples}
\crefname{algo}{Algorithm}{Algorithms}
\crefname{fact}{Fact}{Facts}
\crefname{claim}{Claim}{Claims}
\crefname{appendix}{Appendix}{Appendices}
\crefname{coroll}{Corollary}{Corollaries}
\crefname{figure}{Figure}{Figures}
\crefname{section}{Section}{Sections}

\crefname{thmlisti}{Theorem}{Theorems}
\crefname{lemlisti}{Lemma}{Lemmata}
\crefname{proplisti}{Proposition}{Propositions}
\crefname{asslisti}{Assumption}{Assumptions}
\crefname{deflisti}{Definition}{Definitions}
\crefname{exlisti}{Example}{Examples}
\crefname{algolisti}{Algorithm}{Algorithms}
\crefname{factlisti}{Fact}{Facts}
\crefname{claimlisti}{Claim}{Claims}
\crefname{applisti}{Appendix}{Appendices}
\crefname{MyEnumSeci}{}{}
\crefname{MyEnumSubSeci}{}{}

\makeatletter

\newcommand*{\ie}{%
  \@ifnextchar{,}%
  {\textit{i.e.}}%
  {\textit{i.e.,}\@\xspace}%
}
\newcommand*{\eg}{%
  \@ifnextchar{,}%
  {\textit{e.g.}}%
  {\textit{e.g.,}\@\xspace}%
}
\newcommand*{\etc}{%
  \@ifnextchar{.}%
  {\textit{etc}}%
  {\textit{etc.}\@\xspace}%
}
\newcommand*{\etal}{%
  \@ifnextchar{.}%
  {\textit{et al}}%
  {\textit{et al.}\@\xspace}%
}
\newcommand*{\cf}{%
  \@ifnextchar{.}%
  {\textit{cf}}%
  {\textit{cf.}\@\xspace}%
}
\newcommand*{\aka}{%
  \@ifnextchar{,}%
  {\textit{a.k.a.}}%
  {\textit{a.k.a.}\@\xspace}%
}
\makeatother

\allowdisplaybreaks

\title{PROXIMAL BELLMAN MAPPINGS FOR REINFORCEMENT LEARNING\\
AND THEIR APPLICATION TO ROBUST ADAPTIVE FILTERING
\vspace{-30pt}%
}

\name{}

\address{%
  \begin{minipage}{.7\textwidth}
    \begin{center}
      \textit{Yuki Akiyama\qquad Konstantinos Slavakis}\\[1ex] \small
      Tokyo Institute of Technology, Japan\\
      Department of Information and Communications Engineering\\
      Emails: \texttt{\{akiyama.y.am, slavakis.k.aa\}@m.titech.ac.jp}
    \end{center}
  \end{minipage}
  \vspace{-15pt}
}

\begin{document}
\sloppy
\ninept

\maketitle

\begin{abstract}
  This paper aims at the algorithmic/theoretical core of reinforcement learning (RL) by
  introducing the novel class of proximal Bellman mappings. These mappings are defined in
  reproducing kernel Hilbert spaces (RKHSs), to benefit from the rich approximation properties
  and inner product of RKHSs, they are shown to belong to the powerful Hilbertian family of
  (firmly) nonexpansive mappings, regardless of the values of their discount factors, and
  possess ample degrees of design freedom to even reproduce attributes of the classical Bellman
  mappings and to pave the way for novel RL designs. An approximate policy-iteration scheme is
  built on the proposed class of mappings to solve the problem of selecting online, at every
  time instance, the ``optimal'' exponent $p$ in a $p$-norm loss to combat outliers in linear
  adaptive filtering, without training data and any knowledge on the statistical properties of
  the outliers. Numerical tests on synthetic data showcase the superior performance of the
  proposed framework over several non-RL and kernel-based RL schemes.
\end{abstract}


\section{Introduction}\label{sec:intro}

In reinforcement learning (RL)~\cite{bertsekas2019reinforcement}, an agent takes a
decision/action based on feedback provided by the surrounding environment on the agent's past
actions. RL is a sequential-decision-making framework with the goal of minimizing the long-term
loss/price $Q$ to be paid by the agent for its own decisions. RL has deep roots in dynamic
programming~\cite{bertsekas2019reinforcement, bellman2003dp}, and a far-reaching range of
applications which extend from autonomous navigation, robotics, resource planning, sensor
networks, biomedical imaging, and can reach even to gaming~\cite{bertsekas2019reinforcement}.

This paper aims at the algorithmic/theoretical core of RL by introducing the \textit{novel}\/
class of \textit{proximal Bellman mappings}~\eqref{prox.Bellman}, defined in reproducing kernel
Hilbert spaces (RKHSs), which serve as approximating spaces for the one-step $g$ and long-term
$Q$ losses in RL, and are well known for their rich properties, such as the reproducing
property of their inner product~\cite{aronszajn1950, scholkopf2002learning}. This study stands
as the \textit{first}\/ stepping stone for
\begin{enumerate*}[label=\textbf{(\roman*)}]

\item a simple, flexible and general framework, which can even reproduce attributes of the
  classical Bellman mappings~\eqref{Bellman.maps.standard}, such as their fixed-point sets or
  the mappings themselves (see \cref{prop:special.Bellman}), and for

\item the exciting combination of arguments from RKHSs and nonparametric-approximation
  theory~\cite{Gyorfi:DistrFree:10} with the powerful Hilbertian toolbox of nonexpansive
  mappings~\cite{hb.plc.book}; see \cref{thm:nonexp}.

\end{enumerate*}

Usually in RL, $g$ and $Q$ are considered as points of the Banach space $\mathcal{B}$ of all
(essentially) bounded functions~\cite{bartle.book:95}, equipped with the
$\mathcal{L}_{\infty}$-norm. As such, Bellman mappings operate from $\mathcal{B}$ to
$\mathcal{B}$, they are shown to be contractions~\cite{hb.plc.book} by appropriately
constraining the values of their discount factors (see~\eqref{Bellman.maps.standard}), and
possess thus unique fixed points~\cite{hb.plc.book}. Unfortunately, $\mathcal{B}$ lacks an
inner product by definition. To overcome this inconvenience, the popular strategy in RL is to
assume that $g, Q$ are spanned by a basis of vectors, usually learned from training data, with
a fixed and finite cardinality, which amounts to saying that $g, Q$ belong to a Euclidean vector
space of fixed dimension. These modeling assumptions can be met in almost all currently popular
RL frameworks, from temporal difference (TD)~\cite{bertsekas2019reinforcement} and
least-squares (LS)TD~\cite{lagoudakis2003lspi, regularizedpi:16, xu2007klspi}, to
Bellman-residual (BR) methodologies~\cite{onlinebrloss:16} and kernel-based RL
(KBRL)~\cite{ormoneit2002kernel, ormoneit:autom:02, barreto:nips:11, barreto:nips:12,
  kveton_theocharous_2013, onlinebrloss:16, regularizedpi:16,
  kveton_theocharous_2021}. Notwithstanding, the Bellman mappings introduced
in~\cite{ormoneit2002kernel, ormoneit:autom:02} are still defined on Banach spaces, with no
guarantees that they operate from an RKHS $\mathcal{H}$ to
$\mathcal{H}$. Although~\cite{kerneltd1, wang_principe:spm:21} utilize RKHSs, they do not
discuss Bellman mappings. Proximal mappings have been used in~\cite{proximal-td} only in the
popular context of minimizing loss functions, without any consideration of using proximal
mappings \textit{directly}\/ as Bellman ones.

Unlike typical contraction-based designs in Banach spaces, the novel proximal Bellman
mappings~\eqref{prox.Bellman} are shown to be (firmly) nonexpansive with potentially non-unique
fixed points in RKHSs, \textit{regardless}\/ of the value of their discount factors (see
\cref{thm:nonexp}). This result improves upon the result of the
predecessor~\cite{minh:icassp23} of this work, where the nonexpansivity of the introduced
Bellman mappings was established via appropriate conditions on their discount
factors. Moreover, the benefit of using potentially infinite-dimensional RKHSs comes also from
the freedom of allowing for $Q$-function representations by dynamically changing bases, with
variable and even growing cardinality, to accommodate online-learning scenarios where the basis
vectors are not learned solely from (offline) training data, but may be continuously and
dynamically learned also from streaming (online) test data.

To highlight such online-learning settings, this study considers \textit{robust adaptive
  filtering}~\cite{sayed2011adaptive} as the application domain of the proximal Bellman
mappings~\eqref{prox.Bellman}. The goal is to combat outliers in the classical data-generation
model $y_n = \vectgr{\theta}_*^{\intercal} \vect{x}_n + o_n$, where $n\in\IntegerP$ denotes
discrete time ($\IntegerP$ is the set of all non-negative integers), $\vectgr{\theta}_*$ is the
$L\times 1$ vector whose entries are the system parameters that need to be identified,
$(\vect{x}_n, y_n)$ stands for the input-output pair of available data, where $\vect{x}_n$ is
an $L\times 1$ vector and $y_n$ is real-valued, and $\intercal$ denotes vector/matrix
transposition. Outliers $o_n$ are defined as contaminating data that do not adhere to a nominal
data generation model~\cite{rousseeuw1987}, and are often modeled as random variables (RVs)
with non-Gaussian heavy tailed distributions, e.g., $\alpha$-stable ones~\cite{shao1993signal}.

Since the least-squares (LS) error criterion is notoriously sensitive to
outliers~\cite{rousseeuw1987}, non-LS criteria, such as least mean p-power
(LMP)~\cite{pei1994p-power, xiao1999adaptive, kuruoglu:02, gentile:03, vazquez2012,
  chen2015smoothed, slavakis2021outlier} and maximum correntropy (MC)~\cite{singh.mcc:09}, have
been studied instead of LS ones in robust adaptive filtering. To avoid a lengthy exposition,
this work focuses on the LMP criterion and algorithm~\cite{pei1994p-power}, which, for an
arbitrarily fixed $\vectgr{\theta}_0$, generates estimates
$(\vectgr{\theta}_n)_{n\in\mathbb{N}}$ of $\vectgr{\theta}_*$ as follows:
\begin{equation}
  \vectgr{\theta}_{n+1} \coloneqq \vectgr{\theta}_n + \rho p \lvert e_n\rvert^{p-2}e_n
  \vect{x}_n \,, \label{LMP}
\end{equation}
where $e_n \coloneqq y_n - \vect{x}_n^{\intercal} \vectgr{\theta}_n$, $\rho$ is the learning
rate (step size), and $p$ is a \textit{fixed}\/ user-defined real-valued number within the
interval $[1, 2]$ to ensure that the $p$-norm loss
$\lvert y_n - \vect{x}_n^{\intercal} \vectgr{\theta} \rvert^p$ is a convex function of
$\vectgr{\theta}$~\cite{pei1994p-power}. Notice that if $p = 1$ and $2$, then \eqref{LMP} boils
down to the classical sign-LMS and LMS, respectively~\cite{sayed2011adaptive}. Combination of
adaptive filters with different forgetting factors but with the same fixed
$p$-norm~\cite{vazquez2012}, as well as with different $p$-norms~\cite{chambers1997robust} have
been also considered.

This paper provides an \textit{online}\/ and \textit{data-driven}\/ solution to the problem of
\textit{dynamically}\/ selecting $p$ by using the proposed proximal Bellman mappings,
\textit{without}\/ any prior knowledge on the statistical properties of $o_n$. It is worth
mentioning that this work and its predecessor~\cite{minh:icassp23} are the first attempts in
the literature to apply RL arguments to robust adaptive filtering. \cref{algo} offers an
approximate policy-iteration (API) strategy, where the underlying state space is considered to
be the low-dimensional $\Real^4$, independent of the dimension $L$ of $\vectgr{\theta}_*$,
whereas~\cite{minh:icassp23} uses the high-dimensional $\Real^{2L+1}$. The action space is
considered to be discrete: an action is a value of $p$ taken from a finite grid of the interval
$[1,2]$. Moreover, experience replay~\cite{experiencereplay} (past-data reuse) is introduced,
unlike the classical Bellman operators where information on transition probabilities in a
Markov decision process is required in advance~\cite{bertsekas2019reinforcement}. Note
that~\cite{minh:icassp23} employs rollout for data reuse and
exploration~\cite{bertsekas2019reinforcement}.

Numerical tests on synthetic data showcase the superior performance of the advocated framework
over several non-RL and RL schemes. Due to space limitations, proofs, the convergence analysis
of the proposed algorithm, as well as further RL designs and numerical tests will be reported
elsewhere.

\section{Proximal Bellman Mappings}\label{sec:nonexp.Bellman}

First, some key RL concepts are in order~\cite{bertsekas2019reinforcement}. The state space is
denoted in general by $\mathfrak{S} \subset \Real^D$, with a state vector
$\vect{s} \in \mathfrak{S}$, and the action space by $\mathfrak{A}$, with action
$a\in \mathfrak{A}$. For convenience, the state-action tuple is defined as
$\vect{z} \coloneqq (\vect{s}, a) \in \mathfrak{Z} \coloneqq \mathfrak{S} \times
\mathfrak{A}$. The classical Bellman mappings, \eg, \cite{bellemare:16}:
$\forall (\vect{s}, a) \in \mathfrak{Z}$,
\begin{subequations}\label{Bellman.maps.standard}
  \begin{align}
    (T_{\mu}^{\diamond} Q)(\vect{s}, a)
    & \coloneqq g( \vect{s}, a ) + \alpha
      \mathbb{E}_{\vect{s}^{\prime} \given (\vect{s}, a)} \{ Q(\vect{s}^{\prime},
      \mu(\vect{s}^{\prime})) \}\,, \label{Bellman.standard.mu} \\
    (T^{\diamond} Q)(\vect{s}, a)
    & \coloneqq g( \vect{s}, a ) + \alpha \mathbb{E}_{\vect{s}^{\prime} \given (\vect{s}, a)}
      \{ \inf_{a^{\prime}\in \mathfrak{A} }Q(\vect{s}^{\prime}, a^{\prime})
      \}\,, \label{Bellman.standard}
  \end{align}
\end{subequations}
quantify the total loss (= one-step loss $g$ + expected long-term loss $Q$) that the agent
suffers whenever it takes action $a$ at state $\vect{s}$. Both $g$ and $Q$ map a state-action
tuple $(\vect{s}, a)$ to a real number. In~\eqref{Bellman.maps.standard},
$\mathbb{E}_{\vect{s}^{\prime} \given (\vect{s}, a)}\{\cdot \}$ stands for the conditional
expectation over all possible subsequent states $\vect{s}^{\prime}$ of $\vect{s}$, conditioned
on $(\vect{s}, a)$, and $\alpha$ is the discount factor with typical values in
$(0,1)$. Mapping~\eqref{Bellman.standard.mu} refers to the case where the agent takes actions
according to the stationary policy
$\mu(\cdot): \mathfrak{S} \to \mathfrak{A}: \vect{s} \mapsto \mu(\vect{s})$,
while~\eqref{Bellman.standard} stands as the greedy case of~\eqref{Bellman.standard.mu}. As
explained in \cref{sec:intro}, typically in RL, $T_{\mu}^{\diamond}, T^{\diamond}$ map points
of $\mathcal{B}$ to $\mathcal{B}$. Because of $\alpha\in (0,1)$,
$T_{\mu}^{\diamond}, T^{\diamond}$ are shown to be
contractions~\cite{bertsekas2019reinforcement}, and thus, their fixed-point sets
$\Fix T_{\mu}^{\diamond}$ and $\Fix T^{\diamond}$ are singletons, where
$\Fix T \coloneqq \{ Q\in \mathcal{H} \given TQ = Q\}$ for a mapping
$T: \mathcal{B} \to \mathcal{B}$.

In quest of an inner product, the blanket assumption of this study is that $g, Q$ belong to an
RKHS $\mathcal{H}$, with well-known properties~\cite{aronszajn1950, scholkopf2002learning}, a
reproducing kernel $\kappa(\cdot, \cdot): \mathfrak{Z}\times \mathfrak{Z} \to \Real$, with
$\kappa(\vect{z}, \cdot)\in \mathcal{H}$, $\forall \vect{z}\in \mathfrak{Z}$, and an inner
product which satisfies the reproducing property:
$Q(\vect{z}) = \innerp{Q}{\kappa(\vect{z}, \cdot)}_{\mathcal{H}}$, $\forall Q\in \mathcal{H}$,
$\forall \vect{z}\in \mathfrak{Z}$. Space $\mathcal{H}$ may be infinite dimensional; e.g.,
whenever $\kappa(\cdot, \cdot)$ is the Gaussian kernel~\cite{aronszajn1950,
  scholkopf2002learning}. For compact notations, let
$\varphi(\vect{z}) \coloneqq \kappa(\vect{z},\cdot)$, and
$Q^{\intercal} Q^{\prime} \coloneqq \innerp{Q}{Q^{\prime}}_{\mathcal{H}}$,
$\forall Q, Q^{\prime} \in \mathcal{H}$.

This study introduces the following novel class of \textit{proximal Bellman mappings:} for a
user-defined set of proper and lower-semi-continuous convex functions
$\Set{ f_i\colon \mathcal{H} \to \Real \cup \{+\infty \} }_{i=1}^I$~\cite{hb.plc.book}, define
$T\coloneqq T_{\{f_i\}_{i=1}^I} \colon \mathcal{H} \to \mathcal{H}\colon Q\mapsto TQ \coloneqq
T_{\{f_i\}_{i=1}^I}Q$ as
\begin{align}\label{prox.Bellman}
  TQ & \coloneqq g + \alpha \sum\nolimits_{i=1}^I w_i \prox_{f_i} ( \tfrac{Q-g}{\alpha} ) \,,
\end{align}
where the proximal mapping
$\prox_{f_i}(Q) \coloneqq \argmin_{Q^{\prime}\in \mathcal{H}} f_i(Q^{\prime}) +(1/2)
\norm{Q^\prime - Q}_{\mathcal{H}}^2$~\cite{hb.plc.book}, and coefficients
$\Set{w_i}_{i=1}^I \subset [0,1]$ satisfy $\sum_{i=1}^I w_i = 1$. The user-defined
$\Set{ f_i }_{i=1}^I$ introduce ample degrees of design freedom as the following proposition
demonstrates. Under conditions, mappings \eqref{prox.Bellman} can reproduce
$\Fix T_{\mu}^{\diamond}$ in \cref{prop:case1}, and even replicate~\eqref{Bellman.standard} in
\cref{prop:case2}. Moreover, \eqref{prox.Bellman} open the door to novel RL designs, as
\Cref{prop:case3} exhibits.

\begin{prop}\label{prop:special.Bellman}
  \mbox{}
  \begin{proplist}

  \item\label[proposition]{prop:case1} Consider a stationary policy $\mu(\cdot)$ and define
    $\bar{\varphi}^{\mu} (\vect{z}) \coloneqq \mathbb{E}_{\vect{s}^{\prime} \given \vect{z}}
    \{\varphi(\vect{s}^\prime, \mu(\vect{s}^\prime))\}$, $\forall \vect{z} \in
    \mathfrak{Z}$. Assume that $\bar{\varphi}^{\mu}(\vect{z}) \in \mathcal{H}$,
    $\forall \vect{z}\in \mathfrak{Z}$, and let
    $\bar{h}^{\mu} (\vect{z}) \coloneqq \varphi(\vect{z}) - \alpha
    \bar{\varphi}^{\mu}(\vect{z})$. Assume also that
    $\mathbb{E}_{\vect{s}^{\prime} \given \vect{z}}\{\cdot\}$ interchanges with the inner
    product $\innerp{\cdot}{\cdot}_{\mathcal{H}}$. Further, define
    \begin{align*}
      C^{\mu} \coloneqq \bigcap_{ \vect{z} \in \mathfrak{Z} } \Set*{Q\in\mathcal{H} \given
      \innerp{Q}{\bar{h}^{\mu} (\vect{z})}_{\mathcal{H}} = \mathbb{E}_{\vect{s}^{\prime} \given
      \vect{z}} \{ g( \vect{s}^\prime, \mu(\vect{s}^\prime))\} } \,,
    \end{align*}
    with $C^{\mu}\neq \emptyset$, let $I=1$ in~\eqref{prox.Bellman}, and set
    $f_1 \coloneqq \iota_{C^{\mu}}$, where $\iota_{A}$ stands for the indicator function
    of a set $A\subset \mathcal{H}$, that is, $\iota_{A}(Q) = 0$, if $Q\in A$, and
    $\iota_{A}(Q) = +\infty$, if $Q\notin A$~\cite{hb.plc.book}. Then,
    $\Fix T = \Fix T_{\mu}^{\diamond}$.

  \item\label[proposition]{prop:case2} 

    Define
    $h_Q (\vect{s}, a) \coloneqq \mathbb{E}_{\vect{s}^\prime \given (\vect{s}, a)} \{
    \inf_{a^{\prime} \in \mathfrak{A}} Q(\vect{s}^{\prime}, a^{\prime}) \}$,
    $\forall (\vect{s}, a)\in \mathfrak{S} \times \mathfrak{A}$, and assume that
    $h_Q\in \mathcal{H}$, $\forall Q\in \mathcal{H}$. For $I=1$, the form
    $TQ \coloneqq g + \alpha \prox_{ \iota_{ \{h_Q\} } } ( (Q-g)/\alpha )$
    of~\eqref{prox.Bellman} coincides with~\eqref{Bellman.standard}.

  \item\label[proposition]{prop:case3} For each $i$ in~\eqref{prox.Bellman}, consider a number
    $N^{\textnormal{av}}_i$ of sampling points
    $\{\vect{s}_{ij}^{\textnormal{av}}, a_{ij}^{\textnormal{av}},
    {\vect{s}_{ij}^{\textnormal{av}}}^{\prime} \}_{j=1}^{N^{\textnormal{av}}_i}$, and let
    $\vect{z}_{ij}^{\textnormal{av}} \coloneqq ( \vect{s}_{ij}^{\textnormal{av}},
    a_{ij}^{\textnormal{av}} )$. Consider also a stationary policy $\mu(\cdot)$ and define
    $h^{\mu}_i \coloneqq \varphi(\vect{z}_{i1}^{\textnormal{av}}) - \alpha
    \sum_{j=1}^{N^{\textnormal{av}}_{i}} d_{ij} \varphi(
    {\vect{s}_{ij}^{\textnormal{av}}}^{\prime}, \mu(
    {\vect{s}_{ij}^{\textnormal{av}}}^{\prime}) ) $, for some real-valued averaging weights
    $\{d_{ij}\}$, where $\vect{z}_{i1}^{\textnormal{av}}$ was selected arbitrarily as a
    reference point in the definition of $h^{\mu}_i$. Moreover, by letting
    $g_{ij}^{\mu} \coloneqq g({\vect{s}_{ij}^{\textnormal{av}}}^{\prime}, \mu(
    {\vect{s}_{ij}^{\textnormal{av}}}^{\prime}))$, define the hyperslab
    \begin{align}
      H^{\mu}_i \coloneqq \{ Q\in \mathcal{H} \given\;
      \lvert \innerp{Q}{h^{\mu}_i}_{\mathcal{H}}
      -\sum_{j=1}^{N^{\textnormal{av}}_i} d_{ij} g_{ij}^{\mu} \rvert \leq
      \epsilon_i \} \,,\label{hyperslab}
    \end{align}
    for some user-defined tolerance $\epsilon_i \geq 0$. Let also
    $f_i \coloneqq \iota_{H^{\mu}_i}$. Then, \eqref{prox.Bellman} takes the following form:
    \begin{align}
      TQ {} = {} & Q - \sum_{i=1}^I w_i \frac{\soft_{\alpha\epsilon_i} (
                   \innerp{Q}{h^{\mu}_i}_{\mathcal{H}} - g(\vect{z}_{i1}^{\textnormal{av}})
                   )}{\norm{h^{\mu}_i}_{\mathcal{H}}^2} h^{\mu}_i
                   \,, \label{proximal.hyperslab}
    \end{align}
    where the soft-thresholding function $\soft_{\gamma}(\cdot)$ is defined for some
    $\gamma > 0$ as follows: $\soft_{\gamma}(\xi) = 0$, if $\xi\in [-\gamma, \gamma]$;
    $\soft_{\gamma}(\xi) = \xi - \gamma$, if $\xi\in (\gamma, +\infty)$; and
    $\soft_{\gamma}(\xi) = \xi + \gamma$, if $\xi\in (-\infty, -\gamma)$.

  \end{proplist}

\end{prop}

The following theorem establishes a connection between~\eqref{prox.Bellman} and the Hilbertian
toolbox of nonexpansive mappings~\cite{hb.plc.book}.

\begin{thm}\label{thm:nonexp}
  Mapping \eqref{prox.Bellman} is firmly nonexpansive~\cite{hb.plc.book} in the Hilbert space
  $( \mathcal{H}, \innerp{\cdot}{\cdot}_{\mathcal{H}} )$, regardless of the value of the
  discount factor $\alpha>0$. Whenever
  $\cap_{i=1}^I \argmin_{Q \in \mathcal{H}} f_i(Q) \neq \emptyset$, then
  $\Fix T = g + \alpha \cap_{i=1}^I \argmin_{Q \in \mathcal{H}} f_i(Q)$.
\end{thm}

The online-Bellman-residual (OBR) method~\cite{onlinebrloss:16} shows similarities
with~\eqref{proximal.hyperslab}, since it can be reproduced by~\eqref{proximal.hyperslab} for
the setting where $I=1$, $N_i^{\textnormal{av}} = 1$, $\epsilon_i = 0$, and the scaling factor
$1/\norm{h_i^\mu}_{\mathcal{H}}^2$ does not appear
in~\eqref{proximal.hyperslab}. \cref{thm:nonexp} provides nonexpansivity without constraining
the values of $\alpha$. The predecessor~\cite{minh:icassp23} of this work defined nonexpansive
Bellman mappings in RKHS under assumptions on $\alpha$. The Bellman mappings
in~\cite{ormoneit2002kernel, ormoneit:autom:02, barreto:nips:11, barreto:nips:12,
  kveton_theocharous_2013, kveton_theocharous_2021} share similarities with those
in~\eqref{prox.Bellman}, but they follow the standard RL context, they are viewed as
contractions in $\mathcal{B}$, and no discussion on RKHSs is reported.

\section{Application to Robust Adaptive Filtering}\label{sec:algo}

In this section, mappings~\eqref{prox.Bellman} are applied to the problem of robust adaptive
filtering of \cref{sec:intro}. An RL solution to this problem appeared for the first time in
the literature in the predecessor~\cite{minh:icassp23} of this study. Since online-learning
solutions are of interest, the iteration index of the proposed sequential-decision RL framework
coincides with the time index $n$ of the streaming data $(\vect{x}_n, y_n)_{n\in \IntegerP}$
of~\eqref{LMP}. To this end, the arguments of \cref{sec:nonexp.Bellman} are adequately adapted
to include hereafter the extra time dimension $n$, which will be indicated by the
super-/sub-scripts $[n]$, $(n)$, or $n$ in notations.

\subsection{Defining the state-action space}\label{sec:state.action.space}

The action space $\mathfrak{A}$ is defined as any finite grid of the range $[1,2]$ of the
exponent $p$ in the $\ell_p$-norm. The goal of the proposed RL framework is to generate a
sequence of actions $(a_n = p_n)_{n\in\IntegerP}$ in some ``optimal'' sense. The state space
$\mathfrak{S}$ is assumed to be continuous. Unlike~\cite{minh:icassp23}, where $\mathfrak{S}$
is the high dimensional $\Real^{2L+1}$, this study considers $\mathfrak{S} \coloneqq \Real^4$,
rendering the dimension of $\mathfrak{S}$ independent of $L$ to address the ``curse of
dimensionality'' observed in~\cite{minh:icassp23}. Due to the streaming nature of
$(\vect{x}_n, y_n)_{n\in\IntegerP}$, state vectors
$(\vect{s}_n \coloneqq [ s_1^{(n)}, s_2^{(n)}, s_3^{(n)}, s_4^{(n)}
]^{\intercal})_{n\in\IntegerP}$ are defined inductively by the following heuristic rules:
\begin{subequations}\label{def.states}
  \begin{align}
    s_1^{(n)}
    & \coloneqq \log_{10} (y_n-\vectgr{\theta}_n^\intercal\vect{x}_n)^2
      \,, \label{state1}\\
    s_2^{(n)}
    & \coloneqq \tfrac{1}{M_{\textnormal{av}}} \sum\nolimits_{m=1}^{M_{\textnormal{av}}}
      \log_{10} \frac{ (y_{n-m} - \vectgr{\theta}_n^\intercal
      \vect{x}_{n-m}) ^2}{\norm{\vect{x}_{n-m}}_2^2} \,, \label{state2}\\
    s_3^{(n)}
    & \coloneqq \log_{10} \norm{\vect{x}_n}_2 \,, \label{state3}\\
    s_4^{(n)}
    & \coloneqq \varpi s_4^{(n-1)} + (1 - \varpi) \log_{10} \frac{ \norm{\vectgr{\theta}_{n} -
      \vectgr{\theta}_{n-1}}_2 }{\rho} \,, \label{state4}
  \end{align}
\end{subequations}
where $M_{\textnormal{av}}\in \IntegerPP$, $\varpi \in (0,1)$ are user-defined parameters, and
$\rho$ comes from~\eqref{LMP}. The classical \textit{prior loss}~\cite{sayed2011adaptive} is
used in \eqref{state1}, an $M_{\textnormal{av}}$-length sliding-window sampling average of the
\textit{posterior loss}~\cite{sayed2011adaptive} is provided in \eqref{state2}, normalized by
the norm of the input signal to remove as much as possible its effect on the error, the
instantaneous norm of the input signal appears in \eqref{state3}, and a smoothing
auto-regressive process is used in \eqref{state4} to monitor the consecutive displacement of
the estimates $(\vectgr{\theta}_n)_{n\in\IntegerP}$. The reason for including $\rho$ in
\eqref{state4} is to remove $\rho$'s effect from $s_4^{(n)}$. Owing to \eqref{LMP}, the initial
value $s_4^{(0)}$ in \eqref{state4} is set equal to
$\log_{10}[ (1/\rho) \norm{\vectgr{\theta}_1 - \vectgr{\theta}_0}_2] = \log_{10}{p_0} + (p_0-1)
s_1^{(0)} + s_3^{(0)}$. The $\log_{10}(\cdot)$ function is employed to decrease the dynamic
range of the positive values in \eqref{def.states}.

\subsection{Approximate policy iteration (API)}\label{sec:API}

\begin{algorithm}[t]
  \begin{algorithmic}[1]
    \renewcommand{\algorithmicindent}{1em}

    \State{Arbitrarily initialize $Q_0$, $\mu_0(\cdot)$, and $\vectgr{\theta}_0\in \Real^L$.}

    \While{$n \in \mathbb{N}$}\label{line:iter}

    \State{Data $(\vect{x}_n, y_n)$ become available. Let $\vect{s}_n$ as in
      \eqref{def.states}.}  \State{\textbf{Policy improvement:} Update
      $a_n \coloneqq \mu_n(\vect{s}_n)$ by
      \eqref{policy.improvement}.}\label{algo:policy.improvement}

    \State{Update $\vectgr{\theta}_{n+1}$ by \eqref{LMP}, where $p \coloneqq p_n \coloneqq
      a_n$.}

    \State{Sample
      $\Set{\Set{\vect{z}_{ij}^{\textnormal{av}}[n], {\vect{s}_{ij}^{\textnormal{av}}}^{\prime}
          [n] }_{j=1}^{N_i^{\textnormal{av}}[n]}}_{i=1}^I$, and generate hyperslabs $\Set{
        H_i^{\mu_n} }_{i=1}^I$ (see \cref{sec:API}).}

    \State{\textbf{Policy evaluation:} Update $Q_{n+1}$ by
      \eqref{KM.iteration}.}\label{algo:policy.evaluation}

    \State{Increase $n$ by one, and go to Line \ref{line:iter}.}

    \EndWhile
  \end{algorithmic}

  \caption{Approximate policy iteration for LMP.}\label{algo}

\end{algorithm}

The setting of \cref{prop:case3} is considered here. Since that setting can be viewed as an
approximation of the classical Bellman mappings in~\eqref{Bellman.maps.standard} (see
\Cref{prop:case1,prop:case2}), this section offers the approximate-policy-iteration (API)
\cref{algo} for the problem at hand based on the standard
PI strategy~\cite{bertsekas2019reinforcement}.

With $Q_n$ available, the stationary policy $\mu_n(\cdot)$, which is necessary in
\cref{algo:policy.improvement} of \cref{algo} and for constructing~\eqref{hyperslab} and
\eqref{proximal.hyperslab}, is defined according to the standard greedy
rule~\cite{bertsekas2019reinforcement}
\begin{align}
  \mu_n(\vect{s}) \coloneqq \arg\min\nolimits_{a\in \mathfrak{A}} Q_n( \vect{s}, a)
  \,, \quad \forall \vect{s}\in \mathfrak{S} \,. \label{policy.improvement}
\end{align}
Variations of policy improvement via rollout can be also considered; see for
example~\cite{minh:icassp23}.

To avoid lengthy arguments, only two hyperslabs $\Set{ H_i^{\mu_n} }_{i=1}^{2}$ are utilized
in~\eqref{hyperslab}; that is, $I \coloneqq 2$. Hyperslab $H_1^{\mu_n}$ employs currently and
recently sampled data, while $H_2^{\mu_n}$ employs sampled data from the ``remote past.''
Samples
$\{\vect{z}_{1j}^{\textnormal{av}}[n], {\vect{s}_{1j}^{\textnormal{av}}}^{\prime}
[n]\}_{j=1}^{N_1^{\textnormal{av}}[n]}$ are defined as follows:
$\vect{z}_{11}^{\textnormal{av}}[n] \coloneqq \vect{z}_{n-1} = (\vect{s}_{n-1}, a_{n-1})$ and
${\vect{s}_{11}^{\textnormal{av}}}^{\prime} [n] \coloneqq \vect{s}_n$, and
$\{ (\vect{z}_{1j}^{\textnormal{av}}[n], {\vect{s}_{1j}^{\textnormal{av}}}^{\prime}
[n]) \}_{j = 2}^{N_1^{\textnormal{av}}[n]} \coloneqq \{ (\vect{s}_\tau, a_\tau,
\vect{s}_{\tau+1}) \}_{\tau \in \mathcal{M}[n]}$, where
\begin{align}\label{local sampling}
  \mathcal{M}[n]\coloneqq \{\tau\in\{n-1-N_{\textnormal{w}}, \ldots, n-2\} \given
  \kappa(\vect{z}_{n-1}, \vect{z}_{\tau}) > c\} \,,
\end{align}
for the user-defined sliding-window size $N_{\textnormal{w}}$ and $c > 0$.
The value $g(\vect{z}_{11}^{\textnormal{av}}[n])$ of the one-step loss needed
in~\eqref{proximal.hyperslab} is defined as
\begin{align}
  g(\vect{z}_{11}^{\textnormal{av}}[n]) \coloneqq \tfrac{1}{M_{\textnormal{av}}}
  \sum\nolimits_{m=1}^{M_{\textnormal{av}}} \log_{10} \frac{ (y_{n-m} -
  \vectgr{\theta}_{n}^\intercal\vect{x}_{n-m})^2}{\norm{\vect{x}_{n-m}}_2^2}
  \,. \label{one.step.loss}
\end{align}

Hyperslab $H_2^{\mu_n}$ reuses samples
$\{\vect{z}_{2j}^{\textnormal{av}}[n], {\vect{s}_{2j}^{\textnormal{av}}}^{\prime}
[n]\}_{j=1}^{N_2^{\textnormal{av}}[n]}$ from the ``remote past,'' along the lines
of~\cite{experiencereplay}. Arbitrarily sampling the state-action space to receive feedback
from the surrounding environment (exploration) led to slow adaptation and unstable
performance. This is the reason why samples from the remote past are used here. For
illustration, only a single ``remote-past'' datum, that is,
$N_2^{\textnormal{av}}[n] \coloneqq 1$, is used:
$\vect{z}_{21}^{\textnormal{av}}[n] \coloneqq \vect{z}_{\nu-1} = (\vect{s}_{\nu-1}, a_{\nu-1})$
and ${\vect{s}_{21}^{\textnormal{av}}}^{\prime} [n] \coloneqq \vect{s}_{\nu}$, for some
$\nu < n$. The value $g(\vect{z}_{21}^{\textnormal{av}}[n])$ of the one-step loss is defined as
in~\eqref{one.step.loss}, but with $\nu$ in the place of $n$.

Recalling \Cref{prop:case1,prop:case3}, coefficients
$\vect{d}_i[n] \coloneqq [ d_{i1}[n], \ldots, d_{i N_i^{\textnormal{av}}[n]}[n] ]^{\intercal}$,
which are needed in~\eqref{hyperslab} and~\eqref{proximal.hyperslab}, were introduced so that
$\sum_{j=1}^{N^{\textnormal{av}}_{i}[n]} d_{ij}[n] \varphi(
{\vect{s}_{ij}^{\textnormal{av}}}^{\prime}[n], \mu(
{\vect{s}_{ij}^{\textnormal{av}}}^{\prime}[n]) )$ approximates
$\bar{\varphi}^{\mu} (\vect{z}) = \mathbb{E}_{\vect{s}^{\prime} \given \vect{z}}
\{\varphi(\vect{s}^\prime, \mu(\vect{s}^\prime))\}$. Although there are several ways to
determine $\vect{d}_i[n]$, motivated by the offline-learning context
in~\cite{grunewalder2012modelling}, the following solution to a ridge-regression problem is put
forth here:
\begin{align}
  \vect{d}_i[n]
  & \coloneqq \arg\min_{\vect{d} \in \Real^{N_i^{\textnormal{av}}[n]} }
    \norm*{ \varphi(\vect{z}_{i1}^{\textnormal{av}}[n]) - \vectgr{\Phi}_i^{\textnormal{av}}[n]
    \vect{d} }_{\mathcal{H}}^2 + \sigma_i \norm{ \vect{d} }_2^2 \notag \\
  & = \Set*{ ( {\vectgr{\Phi}_i^{\textnormal{av}}[n]}^\intercal
    \vectgr{\Phi}_i^{\textnormal{av}}[n] + \sigma_i
    \vect{I}_{N_i^{\textnormal{av}}[n]} )^{-1} {\vectgr{\Phi}_i^{\textnormal{av}}[n]}^\intercal
    \varphi(\vect{z}_{i1}^{\textnormal{av}}[n]) } \,, \label{d_k.making}
\end{align}
where
$\vectgr{\Phi}_i^{\textnormal{av}}[n] \coloneqq [\varphi(\vect{z}_{i1}^{\textnormal{av}}[n]),
\ldots, \varphi(\vect{z}_{iN_i^{\textnormal{av}}[n]}^{\textnormal{av}}[n])]$ and
$\sigma_i \geq 0$.

Motivated by the (firm) nonexpansivity of the mappings in~\eqref{prox.Bellman}, established by
\cref{thm:nonexp}, the policy evaluation in \cref{algo:policy.evaluation} of \cref{algo} is
realized by the well-known \textit{Krasnosel'ski\u{i}-Mann}\/ algorithm~\cite{hb.plc.book}:
\begin{align}
  Q_{n+1} \coloneqq (1-\lambda_n) Q_n + \lambda_n T_n(Q_n) \,, \label{KM.iteration}
\end{align}
where $(\lambda_n)_{n\in\IntegerP}$ is a user-defined sequence in $[0,1]$ such that
$\sum_{n\in\IntegerP}\lambda_n ( 1 - \lambda_n ) = +\infty$, and $T_n$ is taken from
\eqref{proximal.hyperslab}.

A direct application of \eqref{proximal.hyperslab} and~\eqref{KM.iteration} may lead to memory
and computational complications, since at each $n$, \eqref{proximal.hyperslab} may add new
kernel functions into the representation of $Q_{n+1}$ via $h_i^{\mu_n}$. This unpleasant
phenomenon is fueled by the potential infinite dimensionality of $\mathcal{H}$; see, for
example, the Gaussian-kernel case~\cite{scholkopf2002learning}. To address this ``curse of
dimensionality,'' random Fourier features (RFF)~\cite{rff} are employed here. Avoiding most of
the details due to space limitations, the feature map $\varphi(\vect{z})$ is approximated by
the following Euclidean vector
\begin{align}
  \tilde{\varphi} (\vect{z}) \coloneqq (\tfrac{2}{D_{\textnormal{F}}})^{1/2} [
  \cos{(\vect{v}_1^{\intercal} \vect{z} + b_1)}, \ldots,
  \cos{\vect{(v}_{D_{\textnormal{F}}}^{\intercal} \vect{z} + b_{D_{\textnormal{F}}})}
  ]^{\intercal} \,, \label{RFF}
\end{align}
with $D_{\textnormal{F}}\in\IntegerPP$ being a user-defined dimension, while
$\{\vect{v}_k\}_{k=1}^{D_{\textnormal{F}}}$ and $\{b_k\}_{k=1}^{D_{\textnormal{F}}}$ are
Gaussian and uniform RVs, respectively.

\section{Numerical Tests}\label{sec:tests}

\begin{figure}[!t]
  \centering
  \subfloat[$\alpha$-stable outliers]{ \includegraphics[ width =
    .23\textwidth]{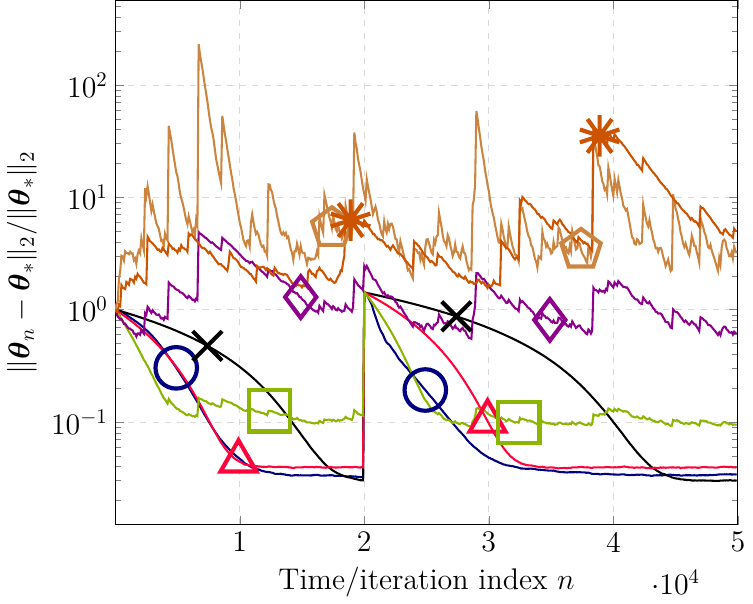} }\label{1-1}
  \subfloat[Sparse outliers]{ \includegraphics[ width =
    .23\textwidth]{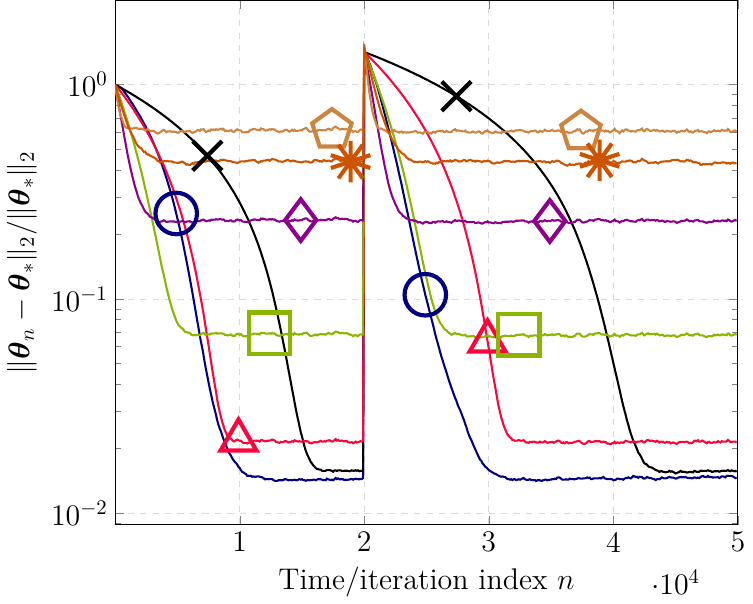} }\label{1-2}
  \caption{\protect\tikz[baseline = -0.5ex]{ \protect\node[mark size = 3pt, color = navy, line
      width = .5pt ] {\protect\pgfuseplotmark{o}}}: \cref{algo} ($\alpha=0.9$). Markers
    \protect\tikz[baseline = -0.5ex]{ \protect\node[mark size = 3pt, color = black, line width
      = .5pt ] {\protect\pgfuseplotmark{x}}}, \protect\tikz[baseline = -0.5ex]{
      \protect\node[mark size = 3pt, color = americanrose, line width = .5pt ]
      {\protect\pgfuseplotmark{triangle}}}, \protect\tikz[baseline = -0.5ex]{
      \protect\node[mark size = 3pt, color = applegreen, line width = .5pt ]
      {\protect\pgfuseplotmark{square}}}, \protect\tikz[baseline = -0.5ex]{ \protect\node[mark
      size = 3pt, color = darkmagenta, line width = .5pt ] {\protect\pgfuseplotmark{diamond}}},
    \protect\tikz[baseline = -0.5ex]{ \protect\node[mark size = 3pt, color = peru, line width =
      .5pt ] {\protect\pgfuseplotmark{pentagon}}} correspond to \eqref{LMP} w/
    $p=1, 1.25, 1.5, 1.75, 2$, respectively. Marker \protect\tikz[baseline = -0.5ex]{
      \protect\node[mark size = 3pt, color = burntorange, line width = .5pt ]
      {\protect\pgfuseplotmark{10-pointed star}}} denotes the algorithm which randomly chooses
    $p$, $\forall n$.}\label{fig:vs.LMP}
\end{figure}

\begin{figure}[!t]
  \centering
  \subfloat[$\alpha$-stable outliers]{ \includegraphics[width =
    .23\textwidth]{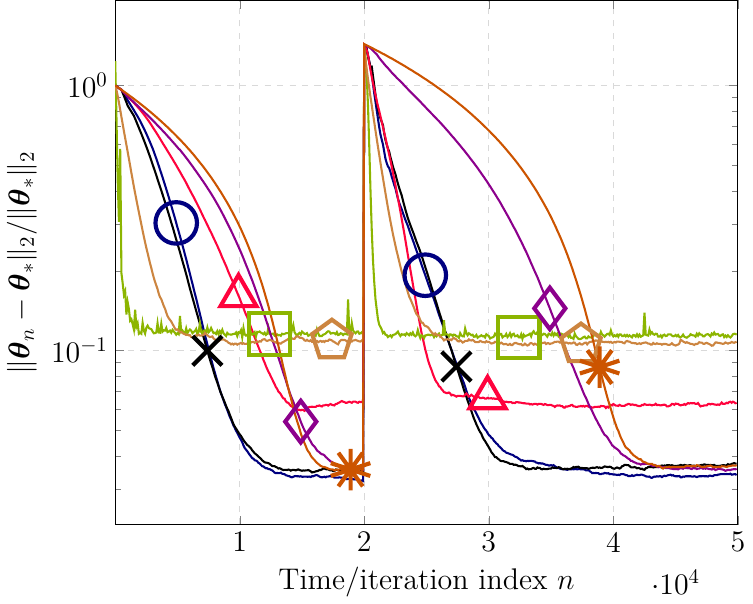}}\label{3-1}
  \subfloat[Sparse outliers]{\includegraphics[width =
    .23\textwidth]{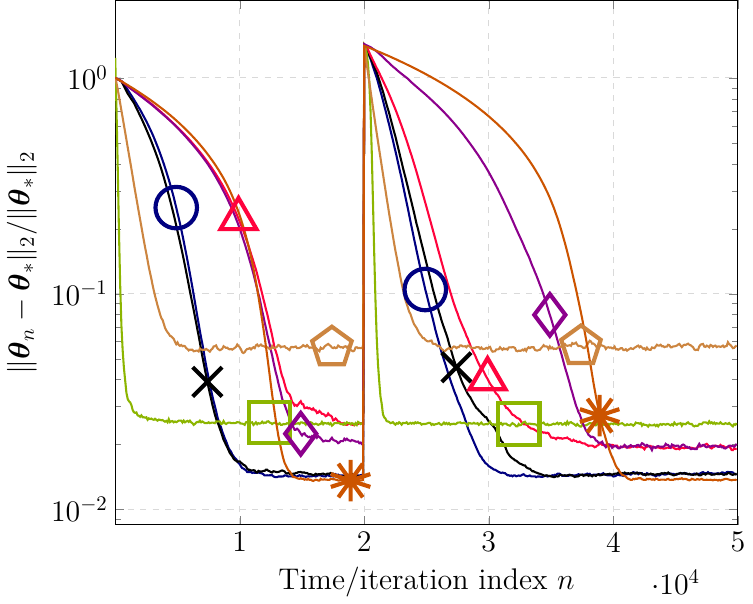}}\label{3-2}
  \caption{\protect\tikz[baseline = -0.5ex]{ \protect\node[mark size = 3pt, color = navy, line
      width = .5pt ] {\protect\pgfuseplotmark{o}}}: \cref{algo}
    ($\alpha=0.9$). \protect\tikz[baseline = -0.5ex]{ \protect\node[mark size = 3pt, color =
      black, line width = .5pt ] {\protect\pgfuseplotmark{x}}}: OBR ($\alpha=0.9$)
    \cite{onlinebrloss:16}. \protect\tikz[baseline = -0.5ex]{ \protect\node[mark size = 3pt,
      color = americanrose, line width = .5pt ] {\protect\pgfuseplotmark{triangle}}}:
    Kernel-based TD(0) ($\alpha = 0.9$)~\cite{kerneltd1}. \protect\tikz[baseline = -0.5ex]{
      \protect\node[mark size = 3pt, color = applegreen, line width = .5pt ]
      {\protect\pgfuseplotmark{square}}}: \cite{vazquez2012}
    ($p=1, \gamma_1 = 0.9, \gamma_2 = 0.99$). \protect\tikz[baseline = -0.5ex]{
      \protect\node[mark size = 3pt, color = darkmagenta, line width = .5pt ]
      {\protect\pgfuseplotmark{diamond}}}: KLSPI ($\alpha=0.9$)~\cite{xu2007klspi}.
    \protect\tikz[baseline = -0.5ex]{ \protect\node[mark size = 3pt, color = peru, line width =
      .5pt ] {\protect\pgfuseplotmark{pentagon}}}: Combination of LMS and
    sign-LMS\cite{chambers1997robust}, \protect\tikz[baseline = -0.5ex]{ \protect\node[mark
      size = 3pt, color = burntorange, line width = .5pt ] {\protect\pgfuseplotmark{10-pointed
          star}}}:~\cite{minh:icassp23}.}\label{fig:vs.TD.KLSPI}
\end{figure}

\begin{figure}[!t]
  \centering \subfloat[$\alpha$-stable outliers]{\includegraphics[width =
    .23\textwidth]{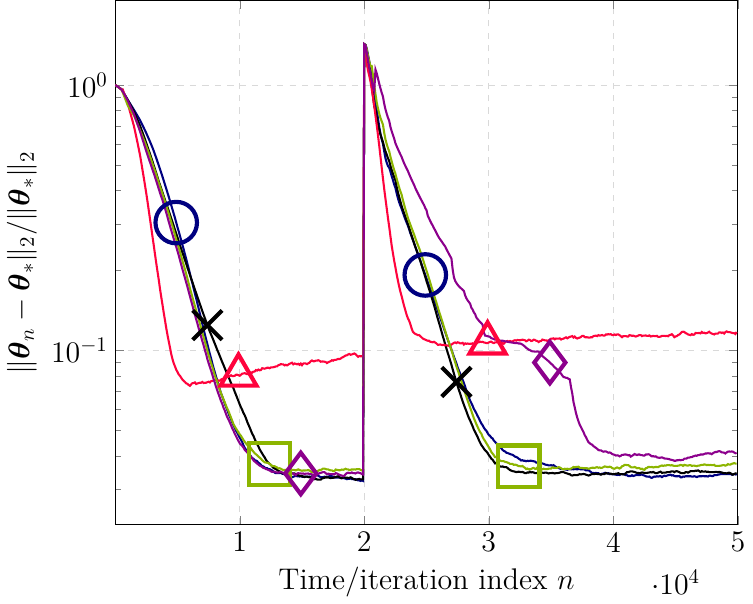}}\label{2-1}
  \subfloat[Sparse outliers]{\includegraphics[width =
    .23\textwidth]{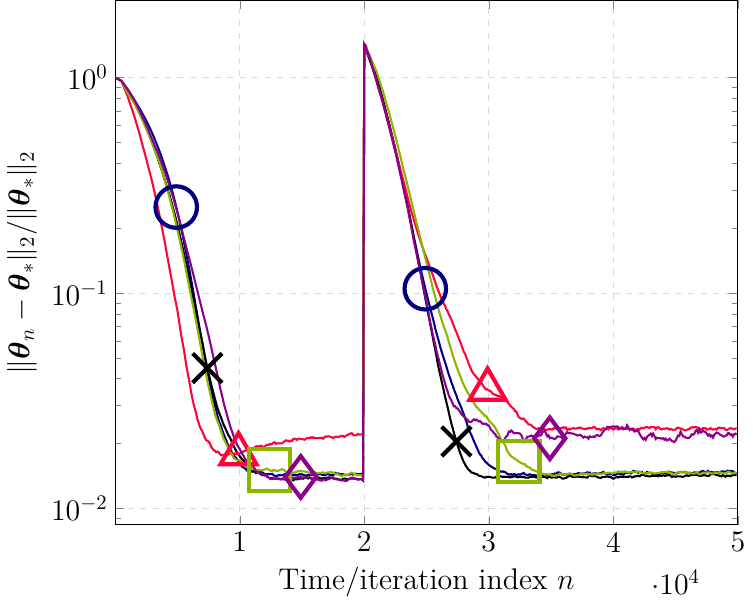}}\label{2-2}
  \caption{\cref{algo} w/ several parameters. \protect\tikz[baseline = -0.5ex]{
      \protect\node[mark size = 3pt, color = navy, line width = .5pt ]
      {\protect\pgfuseplotmark{o}}}: $\alpha=0.9$.  \protect\tikz[baseline = -0.5ex]{
      \protect\node[mark size = 3pt, color = black, line width = .5pt ]
      {\protect\pgfuseplotmark{x}}}: $ \alpha=0.75$.  \protect\tikz[baseline = -0.5ex]{
      \protect\node[mark size = 3pt, color = americanrose, line width = .5pt ]
      {\protect\pgfuseplotmark{triangle}}}: $\alpha=0$.  \protect\tikz[baseline = -0.5ex]{
      \protect\node[mark size = 3pt, color = applegreen, line width = .5pt ]
      {\protect\pgfuseplotmark{square}}}: OBR ($\alpha=0.9$)~\cite{onlinebrloss:16}.
    \protect\tikz[baseline = -0.5ex]{ \protect\node[mark size = 3pt, color = darkmagenta, line
      width = .5pt ] {\protect\pgfuseplotmark{diamond}}}: OBR
    ($\alpha=0.75$)~\cite{onlinebrloss:16}.}\label{fig:vs.params}
\end{figure}

In all tests, the action space $\mathfrak{A} \coloneqq
\{1,1.25,1.5,1.75,2\}$. \Cref{fig:vs.LMP,fig:vs.TD.KLSPI} demonstrate the performance of
\cref{algo} against
\begin{enumerate*}[label=\textbf{(\roman*)}]

\item \eqref{LMP}, where $p\in \mathfrak{A}$ is kept fixed throughout all iterations;
\item \cite{vazquez2012}, which uses a combination of adaptive filters with
  different forgetting factors but with the same fixed $p$-norm;
\item \cite{chambers1997robust}, which uses a combination of LMP \eqref{LMP} ($p=1$ and $p=2$)
  iterations;
\item the kernel-based TD(0)~\cite{kerneltd1} with experience replay and RFF;
\item the online-Bellman-residual (OBR) method~\cite{onlinebrloss:16} with experience replay
  and RFF;
\item the kernel-based (K)LSPI~\cite{xu2007klspi} with experience replay, and
\item the predecessor of this work~\cite{minh:icassp23}.

\end{enumerate*}
Tests were also run to examine the effect of several of \cref{algo}'s parameters on
performance; see \cref{fig:vs.params}. Due to the similarity of OBR
with~\eqref{proximal.hyperslab}, additional realizations of OBR are shown also in
\cref{fig:vs.params}. The metric of performance is the normalized deviation from the desired
$\vectgr{\theta}_*$; see the vertical axes in all figures. The Gaussian
kernel~\cite{scholkopf2002learning} was considered, approximated by RFF as in~\eqref{RFF}. The
dimension $L$ of $\vect{x}_n, \vectgr{\theta}_*$ in \eqref{LMP} is $100$, and the learning rate
$\rho = 10^{-3}$. Both $\vect{x}_n$ and $\vectgr{\theta}_*$ are generated from the Gaussian
distribution $\mathcal{N}(\vect{0}, \vect{I}_{L})$, with $(\vect{x}_n)_{n\in\IntegerP}$ and the
entries of $\vectgr{\theta}_*$ designed to be independent. Moreover,
$M_{\textnormal{av}} \coloneqq 300$ and $\varpi \coloneqq 0.3$ in \eqref{def.states},
$w_1 \coloneqq w_2 \coloneqq 0.5$ and $\epsilon_1 \coloneqq 0, \epsilon_2 \coloneqq 0.05$ in
\eqref{proximal.hyperslab}, while $\lambda_n \coloneqq 0.25$, $\forall n \in \IntegerP$, in
\eqref{KM.iteration}. In~\eqref{d_k.making}, $\sigma_1 \coloneqq 10^{-3}$ and
$\sigma_2 \coloneqq 0$. In~\eqref{local sampling}, $N_{\textnormal{w}} = 10$ and $c=0.95$.

Two types of outliers were considered. First, $\alpha$-stable outliers~\cite{miotto2016pylevy}
are considered, with parameters $\alpha_{\textnormal{stable}} = 1$,
$\beta_{\textnormal{stable}} = 0.5$, $\sigma_{\textnormal{stable}} = 1$, which yield a
considerably heavy-tailed distribution. Second, ``sparse'' outliers are also generated, with
values taken from the interval $[-100, 100]$ via the uniform distribution. Sparse outliers
appear in randomly selected $10\%$ of the total number of time instances in
\Cref{fig:vs.LMP,fig:vs.TD.KLSPI,fig:vs.params}, whereas Gaussian noise with
$\textnormal{SNR} = 30\textnormal{dB}$ appears at every time instance $n$. As it is customary
in adaptive filtering, system $\vectgr{\theta}_*$ is changed randomly at time $20,000$ to test
the tracking ability of \cref{algo}. Each test is repeated independently for $100$ times, and
uniformly averaged curves are reported.

As it can be verified by \Cref{fig:vs.LMP,fig:vs.TD.KLSPI,fig:vs.params}, \cref{algo}
outperforms all competing methods. KLSPI~\cite{xu2007klspi} fails to provide fast convergence
speed. The convergence speed of~\cite{minh:icassp23} is hindered by the high dimensionality of
the adopted state space $\Real^{2L+1}$. TD(0)~\cite{kerneltd1} converges fast, but with a
subpar performance when compared with \cref{algo}. More tests on several other scenarios will
be reported in the journal version of the paper.

\section{Conclusions}\label{sec:conclusions}

The novel class of proximal Bellman mappings was introduced to offer a simple, flexible, and
general framework for reinforcement learning (RL). The proposed framework possesses ample
degrees of design freedom that allows not only for reproducing attributes of the classical
Bellman mappings, widely used in RL, but also to open the door to novel RL designs. The paper
provided also the exciting connection between the advocated proximal Bellman mappings and the
powerful Hilbertian toolbox of nonexpansive and monotone mappings. As a non-trivial application
of the proposed class of mappings, the problem of robust adaptive filtering was considered,
which appears to be addressed under the light of RL for the first time in the literature by
this study and its predecessor. Numerical tests showcase the superior performance of the
proposed design over non-RL and kernel-based RL schemes.

\clearpage

\printbibliography[title = {\normalsize\uppercase{References}}]

@STRING{ieeetsp	= "IEEE Trans.\ Signal Process." }

@STRING{jmlr	= "J.\ Machine Learning Research" }

@STRING{nips	= "Proc.\ NIPS" }

@Article{	  aronszajn1950,
  author	= {Aronszajn, N},
  journal	= {Transactions of the American Mathematical Society},
  pages		= {337--404},
  title		= {Theory of reproducing kernels},
  volume	= {68},
  year		= {1950}
}

@InProceedings{	  barreto:nips:11,
  author	= {Barreto, Andre and Precup, Doina and Pineau, Joelle},
  booktitle	= nips,
  title		= {Reinforcement learning using kernel-based stochastic
		  factorization},
  volume	= {24},
  year		= {2011}
}

@InProceedings{	  barreto:nips:12,
  author	= {Barreto, Andre and Precup, Doina and Pineau, Joelle},
  booktitle	= nips,
  title		= {On-line reinforcement learning using incremental
		  kernel-based stochastic factorization},
  % url		= {https://proceedings.neurips.cc/paper/2012/file/1ecfb463472ec9115b10c292ef8bc986-Paper.pdf},
  volume	= {25},
  year		= {2012},
  % bdsk-url-1	= {https://proceedings.neurips.cc/paper/2012/file/1ecfb463472ec9115b10c292ef8bc986-Paper.pdf}
}

@Book{		  bartle.book:95,
  title		= {The Elements of Integration and Lebesgue Measure},
  % doi		= {https://doi.org/10.1002/9781118164471.fmatter},
  % eprint	= {https://onlinelibrary.wiley.com/doi/pdf/10.1002/9781118164471.fmatter},
  publisher	= {John Wiley \& Sons},
  % url		= {https://onlinelibrary.wiley.com/doi/abs/10.1002/9781118164471.fmatter},
  year		= {1995},
  author	= {Bartle, Robert G},
  % bdsk-url-1	= {https://onlinelibrary.wiley.com/doi/abs/10.1002/9781118164471.fmatter},
  % bdsk-url-2	= {https://doi.org/10.1002/9781118164471.fmatter}
}

@Article{	  bellemare:16,
  author	= {Bellemare, Marc G and Ostrovski, Georg and Guez, Arthur
		  and Thomas, Philip and Munos, Remi},
  chapter	= {Technical Papers: Machine Learning Methods},
  date		= {2016/02/21},
  doi		= {10.1609/aaai.v30i1.10303},
  journal	= {Proc.\ AAAI Conference on Artificial Intelligence},
  journal1	= {AAAI},
  number	= {1},
  title		= {Increasing the action gap: {N}ew operators for
		  reinforcement learning},
  volume	= {30},
  year		= {2016}
}

@Book{		  bellman2003dp,
  author	= {Bellman, Richard Ernest},
  title		= {Dynamic Programming},
  year		= {2003},
  isbn		= {0486428095},
  publisher	= {Dover Publications}
}

@Book{		  bertsekas2019reinforcement,
  title		= {Reinforcement Learning and Optimal Control},
  author	= {Bertsekas, D},
  % isbn		= {9781886529397},
  % url		= {https://books.google.co.jp/books?id=ZlBIyQEACAAJ},
  year		= {2019},
  publisher	= {Athena Scientific}
}

@Article{	  chen2015smoothed,
  author	= {Chen, Badong and Xing, Lei and Wu, Zongze and Liang, Junli
		  and Pr\'{\i}ncipe, Jos\'{e} C and Zheng, Nanning},
  title		= {Smoothed least mean p-power error criterion for adaptive
		  filtering},
  year		= {2015},
  volume	= {40},
  number	= {C},
  issn		= {1051-2004},
  journal	= {Digital Signal Processing},
  month		= may,
  pages		= {154--163},
  numpages	= {10}
}

@InProceedings{	  experiencereplay,
  title		= {Prioritized experience replay},
  author	= {Schaul, T and Quan, J and Antonoglou, I and Silver, D},
  booktitle	= {Proc.\ International Conference on Learning Representations},
  year		= {2016}
}

@Article{	  gentile:03,
  author	= {Gentile, C},
  title		= {The robustness of the p-norm algorithms},
  journal	= {Machine Learning},
  year		= {2003},
  volume	= {53},
  optdoi	= {https://doi.org/10.1023/A:1026319107706},
  pages		= {265--299}
}

@Book{		  hb.plc.book,
  author	= {Bauschke, H H and Combettes, P L},
  title		= {Convex Analysis and Monotone Operator Theory in Hilbert
		  Spaces},
  publisher	= {Springer},
  address	= {New York},
  year		= {2011}
}

@InProceedings{	  kerneltd1,
  author	= {Bae, Jihye and Chhatbar, Pratik and Francis, Joseph T and
		  Sanchez, Justin C and Pr\'{\i}ncipe, Jose C},
  booktitle	= {Proc.\ IEEE EMBS},
  title		= {Reinforcement learning via kernel temporal difference},
  year		= {2011},
  volume	= {},
  number	= {},
  pages		= {5662-5665},
  doi		= {10.1109/IEMBS.2011.6091370}
}

@Article{	  kuruoglu:02,
  title		= {Nonlinear least $\ell_p$-norm filters for nonlinear
		  autoregressive $\alpha$-stable processes},
  journal	= {Digital Signal Processing},
  volume	= {12},
  number	= {1},
  pages		= {119--142},
  year		= {2002},
  optdoi	= {https://doi.org/10.1006/dspr.2001.0416},
  author	= {Kuruo\u{g}lu, Ercan E}
}

@InProceedings{	  kveton_theocharous_2013,
  author	= {Kveton, Branislav and Theocharous, Georgios},
  year		= {2013},
  month		= jun,
  pages		= {569--575},
  title		= {Structured kernel-based reinforcement learning},
  volume	= {27},
  booktitle	= {Proc.\ AAAI Conference on Artificial
		  Intelligence},
  % doi		= {10.1609/aaai.v27i1.8669}
}

@Article{	  kveton_theocharous_2021,
  abstractnote	= {&lt;p&gt; Markov decision processes (MDPs) are an
		  established framework for solving sequential
		  decision-making problems under uncertainty. In this work,
		  we propose a new method for batch-mode reinforcement
		  learning (RL) with continuous state variables. The method
		  is an approximation to kernel-based RL on a set of k
		  representative states. Similarly to kernel-based RL, our
		  solution is a fixed point of a kernelized Bellman operator
		  and can approximate the optimal solution to an arbitrary
		  level of granularity. Unlike kernel-based RL, our method is
		  fast. In particular, our policies can be computed in
		  &lt;em&gt;O&lt;/em&gt;(&lt;em&gt;n&lt;/em&gt;) time, where
		  &lt;em&gt;n&lt;/em&gt; is the number of training examples.
		  The time complexity of kernel-based RL is
		  Ω(&lt;em&gt;n&lt;/em&gt;&lt;sup&gt;2&lt;/sup&gt;). We
		  introduce our method, analyze its convergence, and compare
		  it to existing work. The method is evaluated on two
		  existing control problems with 2 to 4 continuous variables
		  and a new problem with 64 variables. In all cases, we
		  outperform state-of-the-art results and offer simpler
		  solutions. &lt;/p&gt;},
  author	= {Kveton, Branislav and Theocharous, Georgios},
  % doi		= {10.1609/aaai.v26i1.8294},
  journal	= {Proc.\ AAAI Conference on Artificial
		  Intelligence},
  month		= sep,
  number	= {1},
  pages		= {977--983},
  title		= {Kernel-based reinforcement learning on representative
		  states},
  volume	= {26},
  year		= {2021},
  bdsk-url-1	= {https://ojs.aaai.org/index.php/AAAI/article/view/8294},
  bdsk-url-2	= {https://doi.org/10.1609/aaai.v26i1.8294}
}

@Article{	  lagoudakis2003lspi,
  author	= {Lagoudakis, Michail G and Parr, Ronald},
  title		= {Least-squares policy iteration},
  year		= {2003},
  issue_date	= {12/1/2003},
  publisher	= {JMLR.org},
  volume	= {4},
  number	= {},
  % issn		= {1532-4435},
  abstract	= {We propose a new approach to reinforcement learning for
		  control problems which combines value-function
		  approximation with linear architectures and approximate
		  policy iteration. This new approach is motivated by the
		  least-squares temporal-difference learning algorithm (LSTD)
		  for prediction problems, which is known for its efficient
		  use of sample experiences compared to pure
		  temporal-difference algorithms. Heretofore, LSTD has not
		  had a straightforward application to control problems
		  mainly because LSTD learns the state value function of a
		  fixed policy which cannot be used for action selection and
		  control without a model of the underlying process. Our new
		  algorithm, least-squares policy iteration (LSPI), learns
		  the state-action value function which allows for action
		  selection without a model and for incremental policy
		  improvement within a policy-iteration framework. LSPI is a
		  model-free, off-policy method which can use efficiently
		  (and reuse in each iteration) sample experiences collected
		  in any manner. By separating the sample collection method,
		  the choice of the linear approximation architecture, and
		  the solution method, LSPI allows for focused attention on
		  the distinct elements that contribute to practical
		  reinforcement learning. LSPI is tested on the simple task
		  of balancing an inverted pendulum and the harder task of
		  balancing and riding a bicycle to a target location. In
		  both cases, LSPI learns to control the pendulum or the
		  bicycle by merely observing a relatively small number of
		  trials where actions are selected randomly. LSPI is also
		  compared against Q-learning (both with and without
		  experience replay) using the same value function
		  architecture. While LSPI achieves good performance fairly
		  consistently on the difficult bicycle task, Q-learning
		  variants were rarely able to balance for more than a small
		  fraction of the time needed to reach the target location.},
  journal	= {J.\ Mach.\ Learn.\ Res.},
  month		= dec,
  pages		= {1107--1149},
  numpages	= {43}
}

@InProceedings{	  minh:icassp23,
  author	= {Vu, Minh and Akiyama, Yuki and Slavakis, K},
  title		= {Dynamic Selection of p-Norm in Linear Adaptive Filtering
		  via Online Kernel-Based Reinforcement Learning},
  booktitle	= {Proc.\ IEEE ICASSP},
  year		= {2023}
}

@Misc{		  miotto2016pylevy,
  author	= {Miotto, Jos\'{e} Mar\'{i}a },
  title		= {Pylevy},
  year		= {2020},
  howpublished	= {\url{https://github.com/josemiotto/pylevy}}
}

@InProceedings{	  onlinebrloss:16,
  title		= {Online {B}ellman residual and temporal difference
		  algorithms with predictive error guarantees},
  author	= {Sun, Wen and Bagnell, J Andrew},
  year		= {2016},
  optpublisher	= {Carnegie Mellon University},
  pages		= {4213--4217},
  booktitle	= {Proc.\ International Joint Conference on Artificial Intelligence}
}

@Article{	  ormoneit2002kernel,
  title		= {Kernel-based reinforcement learning},
  author	= {Ormoneit, Dirk and Sen, \'{S}aunak},
  journal	= {Machine Learning},
  year		= {2002},
  volume	= {49},
  pages		= {161--178}
}

@Article{	  ormoneit:autom:02,
  author	= {Ormoneit, Dirk and Glynn, Peter},
  title		= {Kernel-based reinforcement learning in average-cost
		  problems},
  journal	= {IEEE Transactions on Automatic Control},
  year		= {2002},
  volume	= {47},
  number	= {10},
  pages		= {1624--1636},
  month		= oct
}

@Article{	  pei1994p-power,
  author	= {Pei, Soo-Chang and Tseng, Chien-Cheng},
  journal	= {IEEE Journal on Selected Areas in Communications},
  title		= {Least mean p-power error criterion for adaptive {FIR}
		  filter},
  year		= {1994},
  volume	= {12},
  number	= {9},
  pages		= {1540--1547},
  doi		= {10.1109/49.339922}
}

@Article{	  regularizedpi:16,
  title		= {Regularized policy iteration with nonparametric function
		  spaces},
  author	= {Farahmand, Amir-Massoud and Ghavamzadeh, Mohammad and
		  Szepesv{\'a}ri, Csaba and Mannor, Shie},
  journal	= jmlr,
  volume	= {17},
  number	= {1},
  pages		= {4809--4874},
  year		= {2016}
}

@InProceedings{	  rff,
  title		= {Random features for large-scale kernel machines},
  author	= {Rahimi, Ali and Recht, Benjamin},
  booktitle	= nips,
  volume	= {20},
  year		= {2007}
}

@Book{		  rousseeuw1987,
  author	= {Rousseeuw, Peter J and Leroy, Annick},
  isbn		= {978-0-47172538-1},
  publisher	= {Wiley},
  optseries	= {Wiley Series in Probability and Statistics},
  title		= {Robust Regression and Outlier Detection},
  year		= {1987}
}

@Book{		  sayed2011adaptive,
  title		= {Adaptive Filters},
  author	= {Sayed, A H},
  optisbn	= {9781118210840},
  optseries	= {IEEE Press},
  url		= {https://books.google.co.jp/books?id=VBaenqIVftUC},
  year		= {2011},
  publisher	= {Wiley}
}

@Book{		  scholkopf2002learning,
  title		= {Learning with Kernels: Support Vector Machines,
		  Regularization, Optimization, and Beyond},
  author	= {Sch\"{o}lkopf, B and Smola, A J},
  % isbn		= {9780262194754},
  % lccn		= {2001095750},
  % series	= {Adaptive computation and machine learning},
  % url		= {https://books.google.co.jp/books?id=y8ORL3DWt4sC},
  year		= {2002},
  publisher	= {MIT Press}
}

@Article{	  shao1993signal,
  author	= {Shao, M and Nikias, C L},
  journal	= {Proc.\ IEEE},
  title		= {Signal processing with fractional lower order moments:
		  {S}table processes and their applications},
  year		= {1993},
  volume	= {81},
  number	= {7},
  pages		= {986--1010},
  doi		= {10.1109/5.231338}
}

@InProceedings{	  singh.mcc:09,
  author	= {Singh, A and Pr\'{i}ncipe, Jos\'{e} C},
  booktitle	= {Proc.\ International Joint Conference on Neural Networks},
  title		= {Using correntropy as a cost function in linear adaptive
		  filters},
  year		= {2009},
  pages		= {2950--2955}
}

@InProceedings{	  slavakis2021outlier,
  author	= {Slavakis, Konstantinos and Yukawa, Masahiro},
  booktitle	= {Proc.\ IEEE ICASSP},
  title		= {Outlier-robust kernel hierarchical-optimization {RLS} on a
		  budget with affine constraints},
  year		= {2021},
  pages		= {5335--5339}
}

@Article{	  vazquez2012,
  author	= {Navia-Vazquez, \'{A}ngel and Arenas-Garcia, Jer\'{o}nimo},
  journal	= ieeetsp,
  title		= {Combination of recursive least p-norm algorithms for
		  robust adaptive filtering in alpha-stable noise},
  year		= {2012},
  volume	= {60},
  number	= {3},
  pages		= {1478--1482},
  doi		= {10.1109/TSP.2011.2176935}
}

@Article{	  wang_principe:spm:21,
  author	= {Wang, Yiwen and Pr\'{\i}ncipe, Jose C},
  journal	= {IEEE Signal Processing Magazine},
  title		= {Reinforcement learning in reproducing kernel {H}ilbert
		  spaces},
  year		= {2021},
  volume	= {38},
  number	= {4},
  pages		= {34--45},
  doi		= {10.1109/MSP.2021.3076309}
}

@Article{	  xiao1999adaptive,
  author	= {Xiao, Yegui and Tadokoro, Y and Shida, K},
  journal	= ieeetsp,
  title		= {Adaptive algorithm based on least mean p-power error
		  criterion for {F}ourier analysis in additive noise},
  year		= {1999},
  volume	= {47},
  number	= {4},
  pages		= {1172--1181},
  doi		= {10.1109/78.752620}
}

@Article{	  xu2007klspi,
  author	= {Xu, Xin and Hu, Dewen and Lu, Xicheng},
  journal	= {IEEE Transactions on Neural Networks},
  title		= {Kernel-based least squares policy iteration for
		  reinforcement learning},
  year		= {2007},
  volume	= {18},
  number	= {4},
  pages		= {973--992},
  % doi		= {10.1109/TNN.2007.899161}
}

@article{chambers1997robust,
  title={A robust mixed-norm adaptive filter algorithm},
  author={Chambers, Jonathon and Avlonitis, Apostolos},
  journal={IEEE Signal Processing Letters},
  volume={4},
  number={2},
  pages={46--48},
  year={1997},
  publisher={IEEE}
}

@article{grunewalder2012modelling,
  title={Modelling transition dynamics in {MDPs} with {RKHS} embeddings},
  author={Grunewalder, Steffen and Lever, Guy and Baldassarre, Luca and Pontil, Massi and Gretton, Arthur},
  journal={arXiv preprint arXiv:1206.4655},
  year={2012}
}

@article{proximal-td,
  author       = {Sridhar Mahadevan and
                  Bo Liu and
                  Philip S. Thomas and
                  William Dabney and
                  Stephen Giguere and
                  Nicholas Jacek and
                  Ian Gemp and
                  Ji Liu},
  title        = {Proximal Reinforcement Learning: {A} New Theory of Sequential Decision
                  Making in Primal-Dual Spaces},
  journal      = {arXiv:1405.6757},
  volume       = {abs/1405.6757},
  year         = {2014},
  url          = {http://arxiv.org/abs/1405.6757},
  eprinttype    = {arXiv},
  eprint       = {1405.6757},
  bibsource    = {dblp computer science bibliography, https://dblp.org}
}

@Book{Gyorfi:DistrFree:10,
  author = {Gy\"{o}rfi, L\'{a}szl\'{o} and Kohler, Michael and Krzy\.{z}ak, Adam and Walk, Harro},
  title = {A Distribution-Free Theory of Nonparametric Regression},
  publisher = {Springer},
  year = {2010},
  address = {New~York}
}

\end{document}